# Design and Experimental Validation of a Software-Defined Radio Access Network Testbed with Slicing Support

K. Koutlia[(1)], R. Ferrus[(1)], E. Coronado[(2)], R. Riggio[(2)], F. Casadevall[(1)], A. Umbert[(1)], J. Pérez-Romero[(1)]

Universitat Politècnica de Catalunya (UPC)[(1)], Fondazione Bruno Kessler (FBK)[(2)]

{katkoutlia, ferrus, ferranc, annau, jorperez}@tsc.upc.edu, {e.coronado, rriggio}@fbk.eu

*Abstract*— Network slicing is a fundamental feature of 5G systems to partition a single network into a number of segregated logical networks, each optimized for a particular type of service, or dedicated to a particular customer or application. The realization of network slicing is particularly challenging in the Radio Access Network (RAN) part, where multiple slices can be multiplexed over the same radio channel and Radio Resource Management (RRM) functions shall be used to split the cell radio resources and achieve the expected behaviour per slice. In this context, this paper describes the key design and implementation aspects of a Software-Defined RAN (SD-RAN) experimental testbed with slicing support. The testbed has been designed consistently with the slicing capabilities and related management framework established by 3GPP in Release 15. The testbed is used to demonstrate the provisioning of RAN slices (e.g. preparation, commissioning and activation phases) and the operation of the implemented RRM functionality for slice-aware admission control and scheduling.

## I. Introduction

5G systems are being designed to support a wider range of applications and business models than previous generations due to the anticipated adoption of 5G technologies in multiple market segments (e.g. automotive, e-health, utilities, smart cities, agriculture, media and entertainment, high-tech manufacturing) and the consolidation of more flexible and cost-efficient service delivery models (e.g. neutral host network providers, Network as a Service, enterprise and private cellular networks). Through the support of network slicing [1], 5G systems are expected to become flexible and versatile network infrastructures where logical networks partitions can be created (i.e. network slices) with appropriate isolation and optimized characteristics to serve a particular purpose or service category (e.g. applications with different access and/or functional requirements) or even individual customers (e.g. enterprises, third party service providers). This is especially relevant for the Radio Access Network (RAN), which is the most resource-demanding (and costliest) part of the mobile network and the most challenged by the support of network slicing [2].

System architecture and functional aspects to support network slicing in 5G Core Network (5GC) and Next Generation RAN (NG-RAN) have already been defined in the first release of the 5G normative specifications approved by 3GPP (e.g. network slice identifiers, procedures and functions for network slice selection, etc.) [3][4]. Moreover, implementation aspects of network slicing in the NG-RAN have been studied from multiple angles, ranging from virtualization techniques and programmable platforms with slice-aware traffic differentiation and protection mechanisms [5]-[7] to algorithms for dynamic resource sharing across slices [8]. In this respect, we analysed in [9] the RAN slicing problem in a multi-cell network in order to show how Radio Resource Management (RRM) functionalities can be used to properly share the radio resources, and we developed in [10], a set of vendor-agnostic configuration descriptors intended to characterize the features, policies and resources to be put in place across the radio protocol layers of a NG-RAN node for the realization of concurrent RAN slices.

On the other hand, management solutions necessary for the exploitation of network slicing capabilities in an automated and business agile manner are at a much more incipient stage, particularly for what concerns the NG-RAN. In this regard, a network slice lifecycle management solution for end-to-end automation across multiple resource domains is proposed in [11], including the RAN domain for completeness, but not addressing it in details. More focused on a 5G RAN, [12] proposes the notion of an on-demand capacity broker that allows a RAN provider to allocate a portion of network capacity for a particular time period, while [13] provides some insight on the need to extend current RAN management frameworks to support network slicing [1] and gives an extensive overview of related use cases. Further progressing on this topic, a functional framework for the management of network slicing for a NG-RAN infrastructure was introduced in [14], detailing the functional components, interfaces and information models that shall be in place, together with a discussion on the complexity of automating the RAN provisioning process. More recently, specifications for a new service-based overall management architecture for 5G systems and network slicing has been concluded by 3GPP as part of Release 15 specifications [15][16]. In this context, building upon the functional framework for RAN slicing management in [14] and consistently with the new service-based management architecture in 3GPP Release 15, this paper describes a Software-Defined RAN (SD-RAN) experimental testbed that allows RAN slices to be automatically provisioned through a RESTful Application Programming Interface (API). Moreover, the testbed implements RRM functions for admission control and scheduling able to treat differently connections belonging to different slices (referred to as *slice-aware* RRM functions in the following). For the characterization of the slicing features at management level, the latest 3GPP Release 15 information models are used as baseline and an extension is proposed to enable a more fine-grained characterisation of the slice-aware RRM policies. The experimental testbed is implemented using open-source RAN distributions (srsLTE [17] and OAI [18]) and the 5G-EmPOWER platform [19]. The testbed is used to experimentally showcase and validate the operation of the slice provisioning phases (e.g. preparation, commissioning and activation of RAN slices) offering the isolation level required, as well as the runtime operation of the implemented slice-aware RRM functionality. Unlike other existing prototypes and Proof of Concepts (PoCs) of RAN slicing features [20][21], the use of the 5G-EmPOWER platform in our tested allows us to come up with a RAN slicing solution that can be ported to different RAN implementations.

The rest of the paper is organized as follows. Section II describes the overall solution framework for RAN slicing management and presents the proposed extension of the 3GPP information models for characterising the RAN slices. Section III describes the experimental platform, discussing the design details of each of the main testbed components and the implemented slice-aware RRM functions. Section IV focuses on showcasing the management and provisioning of RAN slices, while the operational validation and performance assessment of the testbed under a given RAN slicing configuration is addressed in Section V. Finally, concluding remarks and future work are presented in Section VI.

## II. SOLUTION FRAMEWORK FOR AUTOMATED RAN SLICING PROVISIONING

From a service perspective, 3GPP defines a network slice as a particular behaviour delivered by a 5G network. Such behaviour is identified by a Single Network Slice Selection Assistance Information (S-NSSAI) identifier within a Public Land Mobile Network (PLMN). From an implementation perspective, the realization of a network slice is referred to as a Network Slice Instance (NSI). A NSI consists of a set of network function instances and the required resources (e.g. compute, storage and networking resources) that are deployed to serve the traffic associated with one or several S-NSSAIs. A NSI is composed of one or several Network Slice Subnet Instance(s) (NSSI(s)). For example, one NSI can be formed by a NSSI with the RAN functions, denoted in the following as a RAN Slice Instance (RSI), and another NSSI with the 5G core network functions.

Focusing on the RAN part, the implementation of a RSI offers different possibilities on how the NG-RAN infrastructure functions and resources are orchestrated, including how radio spectrum is distributed among RSIs (e.g. RSI-dedicated or shared spectrum). As a general case, let us consider a NG-RAN infrastructure where the base station functions (denoted as gNB for the 5G New Radio [NR] interface) are delivered as a combination of several network functions (e.g. gNB-Distributed Units [gNB-DU] and gNB-Central Units [gNB-CU] hosting different parts of the L3, L2 and L1 radio layer functions) and implemented either as dedicated hardware appliances (i.e. Physical Network Functions (PNFs), after ETSI terminology) or as Virtual Network Functions (VNFs) running on general-purpose hardware, e.g. Network Function Virtualization Infrastructure (NFVI). Under such scenario, a RSI can be realised as a particular configuration of a set of gNB-CU/DU(s) in terms of enabled radio protocol features and radio access capacity guarantees or limitations. The same set of gNB-CU/DU(s) is likely to be shared by several RSIs. On this basis, the support of RAN slicing management capabilities involves different management components as illustrated in Fig. 1. The core functionality consists of a set of management functions, collectively referred to as RAN Slicing Management Function (RSMF) in Fig. 1, in charge of the Lifecycle Management (LCM) of RSIs (e.g. creation, modification and termination of RSIs). To that end, in line with the terminology and service-based management concepts adopted by 3GPP in Release 15 [16], the RSMF exposes a set of management services for the provisioning and monitoring of RSIs (e.g. management services to request the creation of a RSI based on templates, management services to create a measurement job for collecting the performance data of RSIs, etc.). In this respect, the RSMF plays the role of a producer of management services that can be accessed by one or multiple management service consumers. One consumer of the RSMF management services is the network operator (e.g. operator staff that access the RSMF through web-based interfaces or GUIs). Another consumer could be a management system in charge of the LCM of NSIs in which the RSIs are components of the end-to-end slices. In addition, a limited/restricted set of the management capabilities offered by the RSMF might be exposed to the external users (e.g. tenants using the slices) after enforcing the desired exposure governance.

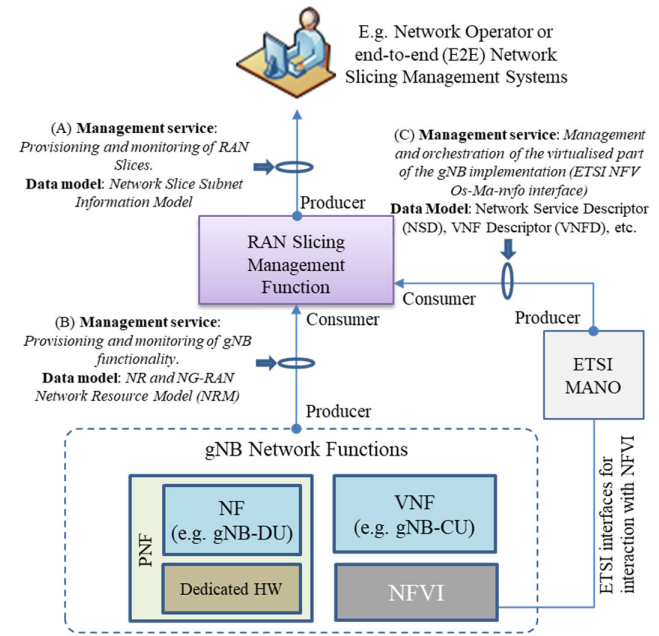

Fig. 1. Functional framework for automated RAN Slicing Management

On the other hand, in order to interact with the underlying infrastructure components and carry out the LCM of the RSIs, the RSMF has to be able to consume the management services provided by the set of PNFs and VNFs hosting the gNB functions. The management of the gNB functions can be realized via standardized interfaces (e.g. the new 3GPP service-based interfaces or legacy 3GPP interfaces such as Itf-N). Through these interfaces, the RSMF would allocate and configure the necessary

L3, L2 and L1 radio access functions and resources (e.g. identifiers, resource reservations) for the creation and operation of RSIs within the gNB functions. Moreover, for the management of NFVI-related aspects of the VNFs hosting part of the gNB functions, the RSMF can consume the management interfaces provided by ETSI NFV Management and Orchestration (MANO) -compliant solutions, such as the Os-Ma-nfvo interface for LCM of network services and VNFs [22].

To support all the interactions across the aforementioned management interfaces, information models to represent the manageable characteristics of the L3, L2 and L1 radio access functions and resources are necessary. In this respect, 3GPP Release 15 specifications define the information models (i.e. managed object classes, attributes, relations) of the components to be managed through the management services tagged as (A) and (B) in Fig. 1. These information models are known as Network Resource Models (NRMs). Specifically, NRM definitions are provided for characterizing a Network Slice Subnet Instance (NSSI)[1] through interface (A), together with NRM definitions for characterizing the gNB functions and its supported slicing features through interface (B). For the sake of having a comprehensive view of the scope of the NRM definitions, Table 1 outlines the main attributes in both NRMs that directly cope with the slicing features.

From Table 1, it can be noted that the slicing modelling established by 3GPP basically intends to define a minimum set of high-level attributes for network slicing management. This approach brings high flexibility as very few constraints are imposed on the specific configurations, though only allows for a coarse-grained management capability, especially when it comes to the configuration of the radio resources in the gNB for RAN slicing. Therefore, building upon these high-level 3GPP models, a more fine-grained management of the RAN slices necessarily requires the extension or addition of new attributes for a more precise characterization of the behavior of a RAN slice. In this regard, focusing on the problem of how to split the radio resources between multiple slices, the solution proposed in this paper develops the semantics of the *RRMPolicy* attribute of the gNB NRM, which is implementation dependent, to convey a more detailed configuration of the RAN slices. This is carried out by leveraging our previous work [10], in which a set of comprehensive configuration descriptors was proposed to parametrize the features, policies and resources put in place across the L1, L2 and L3 radio protocol layers distributed across the set of PNFs and VNFs that jointly provide the full radio access functions.

An illustration of the formulation of the *RRMPolicy* attribute based on the proposed descriptors is depicted in Fig. 2. It consists of three configuration descriptors, referred as *L3*, *L2* and *L1 slice descriptors*, which are used to characterize the operation of the underlying radio protocol layers for the realization of the RAN slice. A brief description of these descriptors that is sufficient to base the subsequent design and implementation aspects is provided below. Further details on the different parameters included in the descriptors, together with the rationale behind, can be found in [10].

With regard to the *L3 slice descriptor*, L3 comprises the Radio Resource Configuration (RRC) protocol and RRM functions such as Radio Bearer Control (RBC), Radio Admission Control (RAC) and Connection Mobility Control (CMC) for the activation and maintenance of Radio Bearers (RB), which are the data transfer services delivered by the radio protocol stack. For each UE, one or more user plane RBs, denoted as Data RBs (DRBs), can be established per Protocol Data Unit (PDU) session, which defines the connectivity service provided by 5GC [3]. A *L3 slice descriptor* is necessary to specify the capacity allocation for the RAN slice (e.g. number and characteristics of the DRBs that can be simultaneously established), the RRM policies that govern the operation of the slice (e.g. DRB configuration policies) and the capability set of the RRC protocol in use (e.g. application type specific RRC messages).

| NSSI NRM (SliceProfile class) ||
|---|---|
| Attribute | Explanation |
| SliceProfileId | A unique identifier of the slice profile. |
| SNSSAI | Set of supported S-NSSAI(s) in the NSSI. Each S-NSSAI is comprised of a SST (Slice/Service type) and an optional SD (Slice Differentiator) field. |
| PLMNId | Set of PLMN(s) associated with the NSSI. |
| PerfReq | It specifies the requirements to the NSSI in terms of the scenarios defined in the TS 22.261, such as experienced data rate, and area traffic capacity (density) of UE density. Limitation of the attribute values is not addressed. |
| maxNumberofUEs | It specifies the maximum number of UEs that may simultaneously access the NSSI. |
| coverageAreaTAList | List of tracking area(s) where the NSSI can be selected. |
| Latency | It specifies the packet transmission latency (millisecond) through the RAN, CN, and TN part of 5G network |
| UEMobilityLevel | It specifies the mobility level of a UE accessing the NSSI. Allowed values: stationary, nomadic, restricted mobility, and fully mobility. |
| ResourceSharingLevel | It specifies whether the resources to be allocated to the network slice instance may be shared with another network slice instance(s). Allowed values: shared, non-shared. |
| NR and NG-RAN (gNB) NRM ||
| Attribute | Explanation |
| SNSSAI | Set of supported S-NSSAI(s). |
| RRMPolicy | Represents the RRM policy, which includes guidance for split of radio resources between multiple slices the cell supports. The RRM policy is implementation dependent. |

Table 1. 3GPP defined attributes for network slicing configuration

As to the *L2 slice descriptor*, L2 comprises a Medium Access Control (MAC) sub-layer for multiplexing and scheduling the packet transmissions of the DRBs over a set of transport channels exposed by L1. Moreover, L2 embeds a number of processing functions configurable on a per-DRB basis for e.g. segmentation, Automatic Repeat reQuest (ARQ) retransmissions, compression and ciphering (i.e. Radio Link Control [RLC] and Packet Data Convergence Protocol [PDCP]). In the NR specifications, an additional L2 sub-layer named Service Data Adaptation Protocol (SDAP) is included to map the DRBs and the traffic flows managed by the 5GC, referred to as QoS Flows [3] SSI. Considering that the current MAC operation is based on individual UE and DRB specific QoS profiles, a *L2 slice*

---

[1] A NSSI is the central building block of a Network Slice Instance [NSI], that is, a NSI is always composed of one or several NSSI(s) [15]. Note that a RAN Slice Instance (RSI) as defined in this paper is a particular realization of a NSSI for the NG-RAN functionality.

*descriptor* is necessary to define the packet scheduling behavior to be enforced on the traffic aggregate of DRBs of the same slice and to specify the capability set of the applicable L2 sub-layers processing functions.

With regard to the *L1 slice descriptor*, L1 provides L2 with transfer services in the form of transport channels, which define the way in which data is transferred (e.g. Transmission Time Interval [TTI], channel coding). L1 also establishes the corresponding radio resource structure of the cell radio resources (e.g. waveform characteristics and time/frequency domain resource structure). Considering that a RAN slice may require specific L1 transfer service capabilities (e.g. low latency shared transport channel) and/or specific radio resource allocation of the cell radio resources, a *L1 slice descriptor* is needed to specify both aspects.

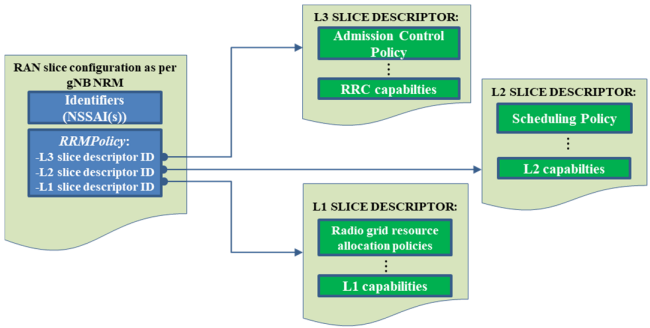

Fig. 2. Implementation of the *RRMPolicy* attribute semantics to represent a set of descriptors for the configuration of the L3, L2 and L1 radio access functions for the realization of a RSI

## III. SOFTWARE-DEFINED RAN (SD-RAN) EXPERIMENTAL TESTBED

A high-level view of the experimental testbed for showcasing and validating the solution framework for RAN slicing management presented in Section II is depicted in Fig. 3. The testbed includes a disaggregated RAN implementation, in which some control-plane functions in the RAN (e.g. slice-aware/multi-cell admission control) are centralized in the form of a SD-RAN controller. Together with the RAN control-plane functionality, the SD-RAN Controller also integrates the RSMF function, which is responsible for LCM of the RSIs. The SD-RAN Controller has been implemented as an extension of the 5G-EmPOWER Operating System (OS), which is an open-source experimental platform for 5G-service development and testing [19]. As illustrated in Fig. 3, the SD-RAN Controller provides a RESTful API for the provisioning of slices and the so-called 5G-EmPOWER Northbound API for running control and management applications on top of the controller, such as the multi-cell/slice-aware admission control and multi-cell/slice-aware scheduling coordination functions explained later on.

The interaction of the SD-RAN Controller with the distributed functions of the RAN is performed through the 5G-EmPOWER Agent, which is a functionality embedded within the access nodes. The testbed used for the prototype is based on a research-oriented open-source SDR implementation of a LTE eNB (i.e. srsLTE [17]) and Ettus Research Universal Software Radio Peripherals (USRPs) b210, as depicted in Fig. 3. Notice that the support of specific radio front-ends is up to the implementer of the network stack (i.e. in this case up srsLTE), and independent from the 5G-EmPOWER capabilities, which are compatible with any radio front-end supported by the stack. It is important to highlight that although the 5G-EmPOWER OS is able to support 4G and 5G networks, the validation of the prototype is at the moment limited to 4G since no open-source 5G stacks are currently available for experimentation. Accordingly, the eNBs are connected in the testbed to an open-source Evolved Packet Core (EPC) (e.g. OAI [18], NextEPC [23] implementations to, jointly with the RAN, provide IP connectivity services to a number of off-the-shelf User Equipment (UE) terminals used for testing. The EPC, the eNBs and the SD-RAN controller are deployed on an Intel NUC with an i5 Intel processor and 16 GB of RAM memory running Ubuntu 18.04.1. However, alternatively, each of these components can be also deployed in independent off-the-shelf laptops equipped with the same processor family and at least 4GB of RAM. More details on each component are given in the following.

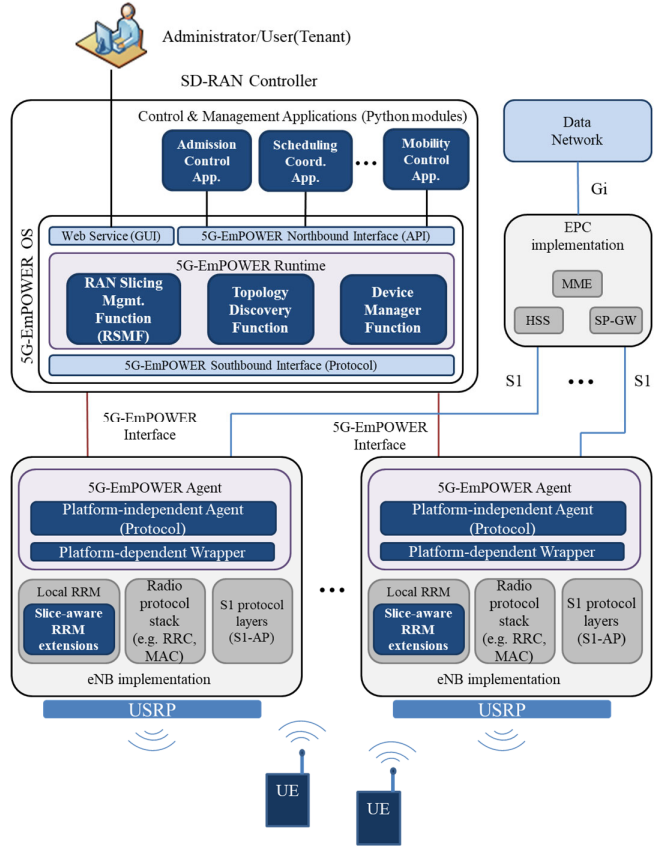

Fig. 3. SD-RAN Experimental Testbed with RAN Slicing Management features

### A. The 5G-EmPOWER Operating System

The 5G-EmPOWER OS provides a framework for managing heterogeneous RAN nodes (e.g. 3GPP RAN nodes, Wi-Fi access points) based on the OpenEmpower protocol, together with a collection of built-in functionalities, services and APIs that can be used to program control and management applications. The 5G-EmPOWER OS is implemented in Python using the Tornado Web Server as web framework.

The internal implementation of the OS follows a modular architecture. As a matter of fact, except for the logging subsystem (which must be available before any other module is loaded), every task supported by the 5G-EmPOWER OS is implemented as a plug-in (i.e. a Python module) that can be loaded at runtime. Modules can be built-in and launched at bootstrap time or started and stopped at runtime. Each module consists of a Manifest file containing the module meta-data (version, dependencies, etc.), and one or more Python scripts. Developers are free to decide how their network control and management applications are deployed. For example, the developed applications can implement control capabilities per RAN slice or for the overall operation of the RAN. Moreover, an application can consist of a single or several modules. This approach is similar, in principle, to the Network Function Virtualization (NFV) paradigm, where complex services can be deployed by combining several VNFs. Likewise, in the 5G-EmPOWER OS, complex network management applications can be designed by deploying and combining different modules. This allows developers to dynamically set up a network monitoring application to perform a site survey or to roll out new features at runtime by selecting them from an "app store". In the implemented testbed, the 5G-EmPOWER OS includes three core modules, namely:

- Device Manager Function. Tracks the eNBs active in the RAN. This includes their IP address and their identifier, i.e. the eNB ID, the last seen date, and a list containing the capabilities of the eNB (e.g. DL/UL EARFCN[2]). The device manager exposes an API allowing applications to receive events when new eNBs join or leave the RAN.
- Topology Discovery Function. This function is implemented as a collection of modules that allow the controller to collect static (with long term dynamics), as well as dynamic (short term) information of the network. The first type comprises information about how RAN nodes are interconnected (i.e. how the eNBs are interconnected with each other through, for example, X2 interfaces) in order to build a logical connection map of the network. This map is updated by the SD-RAN controller upon the reception of events raised by the Agent when links are added or removed, case that does not occur in a frequent basis. The second type is related to periodic measurements from the eNBs and the UEs (e.g. RSRP/RSRQ). Notice that the notification period of these messages can be adjusted by the SD-RAN controller depending on the network load. In addition, the eNB can aggregate in a single message the information of all the UEs attached to it, therefore not affecting negatively the scalability of the system. Let us notice that this long term and short term information can be used by control and management applications running on top of the SD-RAN Controller. Moreover, although not exploited in this work, the topology discovery modules may be fed with information from external sources (e.g. spectrum databases).
- RAN Slicing Management Function (RSMF). It is responsible for the entire slice lifecycle management, from provisioning to decommissioning. This module supports the instantiation of RAN slices whose operation is dictated by the *L3*, *L2* and *L1 slice descriptors* defined in Section II. In particular, Table 2 depicts the template proposed for the specification of a RSI and the configuration of the RRM policy for specifying the behavior of both L3 through admission control and L2 through scheduling policies. Further details on the semantics of the RRM policy L3 and L2 attributes are given later on.

In addition to the RSMF, the support of RAN slicing features requires the deployment of control applications to cope with the slice-aware/multi-cell RRM functions necessary for the proper handling of the radio resources within and between slices. These control applications, whose functionality is explained below in a separate subsection, interact with the 5G-EmPOWER OS through a set of APIs (northbound API in Fig. 3) designed with the express goal of shielding developers from the implementation details of the underlying wireless technology. In line with the implementation approach followed for the 5G-EmPOWER OS, the northbound APIs are provided as Python libraries so that the writing of new applications is facilitated. This design choice allows programmers to leverage a high-level declarative API, while being able to use any Python construct, such as threads, timers, sockets, etc.

Finally, the 5G-EmPOWER OS exposes a set of management services (e.g. slice creation, network inventory, monitoring) to the platform administrator (e.g. operator role for the RAN), as well as to other potential users of the system (e.g. tenant role in multi-tenant RAN). This allows for an exploitation model where the platform administrator would own and operate both the radio infrastructure and the SD-RAN Controller, while the tenants (e.g. 3rd party service providers, verticals) would be consumers of the exposed management services (i.e. a restricted set of the management services authorized by the platform administrator), which could be integrated within their own management systems. The management services are offered through a RESTful API implemented by a Web Service module. This functionality is split into two submodules: the REST server and the front-end Graphical User Interface (GUI). The benefit of this approach is that the 5G-EmPOWER OS is not GUI dependent and any client that can consume a REST service can interact with the OS.

*B. The 5G-EmPOWER Agent*

The 5G-EmPOWER Agent is in charge of managing the LTE user plane. An eNB integrating the 5G-EmPOWER Agent within its subsystem can interact with the 5G-EmPOWER OS. The architecture of the 5G-EmPOWER Agent is composed of two parts written in C++: the platform independent 5G-EmPOWER Agent itself and the platform dependent Wrapper. The agent consists of: (i) a protocol parser responsible for serializing and de-serializing the OpenEmpower messages; and (ii) two managers, one for single/scheduled events and one for triggered events. The types of events supported by the Agent are described in the next subsection. Finally, the wrapper is responsible for translating OpenEmpower messages into commands for the LTE stack. Fig. 3 sketches the structure of the 5G-EmPOWER Agent.

---

[2] Downlink/Uplink E-UTRA Absolute Radio Frequency Channel Number

The Agent and the OpenEmpower protocol are generic and can be applied to any general eNB. However, the Wrapper must be written for a specific eNB implementation since it defines a set of operations that an eNB must support in order to be part of a 5G-EmPOWER-managed network. Such operations include, for example, getting/setting certain parameters from the LTE access stratum, triggering UE measurements reports, rising UE attach/detach events, issuing commands (e.g. perform a handover), and reconfiguring a certain access stratum policy. All these operations are invoked by the 5G-EmPOWER OS through the OpenEmpower Protocol. Notice that the implementation of each of these features is responsibility of the eNB vendor and platform dependent, which makes it unrelated from the capabilities offered by the 5G-EmPOWER system.

The Wrapper is structured in as many submodules as the layers in the LTE Access Stratum plus an additional module for the RRC functions. Finally, it is important to highlight that the implementation of each of these submodules is responsibility of the eNB vendor and platform dependent, which makes it unrelated from the capabilities offered by the 5G-EmPOWER OS. At the time of writing, there is available a reference implementation of the 5G-EmPOWER Agent for OpenWRT-based Wi-Fi APs, for LTE small cells based on the srsLTE stack [17] and for a few commercial 4G/5G eNBs.

| Attributes | | Description |
|---|---|---|
| Identifiers | | **RAN Slice ID (RSI ID)**: Identifies the RSI. |
| | | **PLMNList:** Set of PLMN(s) served through the RSI. |
| | | **SNSSAIList**: Set of S-NSSAI(s) per PLMN served through the RSI. |
| | | **CellList:** Set of cells where the **NSSAIList** can be selected. |
| RRMPolicy | L3 Descriptor | **Aggregated radio load** [units: %][(QCI, ARP)[1] pairs][cell-level, slice-level][Minimum, Maximum] |
| | | **Aggregated bit rate** [units: b/s] [(QCI, ARP) pairs][cell-level, slice-level][Minimum, Maximum] |
| | | **Number of DRBs** [(QCI, ARP) pairs][cell-level, slice-level][Minimum, Maximum] |
| | | **Number of UEs** [cell-level, slice-level][Minimum, Maximum] |
| | | **Averaging Window:** It represents the duration over which the radio load and bit rate shall be calculated. |
| | L2 Descriptor | **Inter-slice scheduling:** [algorithm], optionally [resource units per slice: %] |
| | | **Intra-slice scheduling:** [algorithm] |

[1] QCI (QoS Class Identifier) and ARP (Allocation and Retention Priority) are the two mandatory atributes in the QoS model for radio bearers.

Table 2. Template with the descriptors for the specification of a RAN Slice

## C. The OpenEmpower Protocol

As stated in the previous section, an agent is introduced at the eNB to implement the management actions defined by the OS layer. Communication between the agent and the OS layer happens over the OpenEmpower protocol. The 5G-EmPOWER OS provides a reference implementation of the OpenEmpower protocol, however implementations for other SDN platforms are also possible, e.g. ONOS or OpenDayLight. The OpenEmpower protocol allows remote management of RAN elements, while it makes no assumption about the type of RAN element, i.e. it can be used on Wi-Fi APs, LTE eNBs, or 5G gNBs.

The protocol is built on three major events or message types: single-event, scheduled-event, and triggered-event. Their meaning is the following:

- Single Events. These are simple standalone events requested by the OS plane and notified back immediately by the agent. No additional logic is bound to such message and the OS decides the time to issue the next event. Examples include RAN element capabilities requests or handover requests.
- Scheduled Events. These are events initiated by the OS plane and then executed periodically by the agent. Examples include the Physical Resource Block (PRB) utilization requests, which require the agent to periodically send a PRB utilization report to the OS plane.
- Triggered Events. These events enable/disable a certain functionality at the agent. They specify a condition that, when verified, triggers a message from the agent to the 5G-EmPOWER OS. Examples include the RRC measurements requests.

All OpenEmpower protocol messages start with a common header that specifies the protocol version, the event type, the message length, the RAN element ID (e.g. eNB ID) and the cell ID, the transaction ID, and a 32-bits sequence number. The counter associated to the sequence number is independent for the connection between each eNB and the 5G-EmPOWER OS, and is incremented by one every time a message is generated by either an agent at the eNBs or the OS plane. The transaction ID is a 32-bits token associated with a certain request. Replies must use the same ID as in the request in order to facilitate pairing. This is necessary because all the communications using the OpenEmpower protocol are asynchronous.

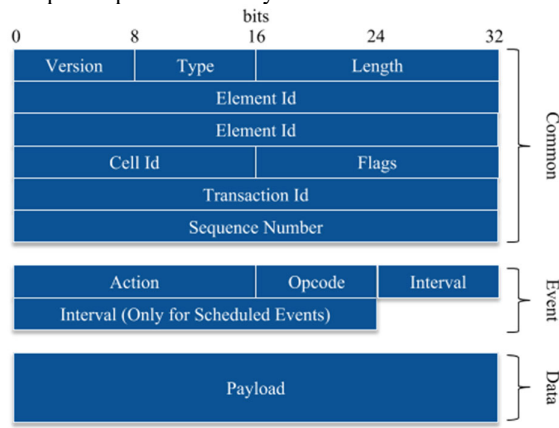

Fig. 4. The OpenEmpower message structure

The common header is followed by one of the three possible events headers. Each event header specifies the type of action, an operation code (opcode), and (in the case of a scheduled event), the event scheduling period. The opcode value depends on the particular type of action and can be used to indicate both error/success conditions or the type of operation (create, retrieve, update, or delete). Finally, after the event header, we can find the body of the message itself, which differs from action to action. Figure 4 sketches the structure of an OpenEmpower message.

*D. Slice-aware RRM functions*

The enforcement of the RRM policies defined to characterize the expected operation of a RSI (see *RRMPolicy* attributes in Table 2) requires the implementation within the testbed of a slice-aware/multi-cell scope RRM functions for Admission Control (AC) and for scheduling.

More specifically, the AC function already supported in the eNB (srsLTE implementation) is extended in order to be compliant with the proposed descriptors and, as a whole, to oversee the capacity allocation and utilization per slice and across all the cells and decide on the acceptance or denial of the DRBs. The operation of the AC is dictated by the "Admission Control Policy" L3 attribute within the RSI template, whose detailed semantics have been presented in Table 2. As shown in the table, the capacity allocation is defined as a combination of the aggregated radio load, the aggregated bit rate and the number of established DRBs, each of them specifiable at a granularity down to particular pairs of QCI and ARP values and with the possibility to define them per cell and/or per slice (multi-cell) level. In addition, the *L3 slice descriptor* also includes the number of connected UEs. All these parameters can be used to establish minimum capacity guarantees and/or capacity limitations.

The implementation of the AC function is carried out by extending the legacy RRM functions of the eNB with slice-aware AC capabilities at cell level and by a control application (*Admission Control Application* depicted in Fig. 3) executed at the SD-RAN Controller with multi-cell AC capabilities. This split of the AC function between the eNB and the SD-RAN controller reduces the amount of information exchanged between the eNB and the SD-RAN Controller in contrast to solutions in which the AC function could be entirely embedded in the eNB (i.e. fully distributed implementation) or entirely embedded in the SD-RAN controller (i.e. fully centralized implementation). Notice that in a fully centralized solution with AC capabilities only at the SD-RAN controller, much stress would be put on the eNB monitoring signaling to have a very accurate view of the radio load in each cell. On the other hand, on a fully distributed solution, with AC decisions only local to eNBs, a significant amount of information should be conveyed among eNBs in order to allow each of them to make decisions with a multi-cell/slice-aware scope.

One of the challenges met during the implementation of the distributed AC function has been the optimization of the message exchange delay between the centralized AC module and the eNB to support centralized decisions without causing a RRC timer expiration on the connected UEs.

On the other hand, in line with approaches such as the Network Virtualization Substrate (NVS) concept presented in [24] the scheduling function within the eNB implementation has been extended to cope with the resource allocation policies intended to regulate the distribution of the radio resources of a cell (i.e. PRBs) among the groups of UEs associated with different RAN slices. In particular, the portion of the radio resources to be assigned to a particular slice is dictated by the choice of the *L2 slice descriptor* (see Table 2), which is specified at SD-RAN Controller level and configured at the managed eNBs through the 5G-EmPOWER Agent. In this regard, the scheduling policy is defined through two attributes:

- Inter-slice scheduling. It defines the treatment that a slice is given with respect to the others. It is specified in terms of the algorithm used and, optionally, the percentage of resources (e.g. percentage of PRBs) allocated to each slice per TTI. The algorithm selected for inter-slice scheduling may be of various types. For example, when setting a classic Round Robin (RR) algorithm the slices are assigned the same portion of the available resources; however, this scheduler is not QoS-based and does not enforce any particular distribution of the resources. Conversely, by specifying a Weighted Round Robin (WRR) and the percentage parameter, it is possible to enforce the desired distribution of resources in terms of PRBs.
- Intra-slice scheduling. It defines how the UEs within a given/particular slice are scheduled. It is specified only in terms of the scheduling algorithm used in the particular slice. Different slices can be configured with different algorithms. The sort of algorithms that can be selected currently are the common scheduling methods used in non-sliced settings, such as Round Robin (RR), Proportional Fair (PF) and max C/I [25]. These algorithms allocate (to the different UEs) the resources assigned to each slice by the SD-RAN controller considering the QoS characteristics of the established DRBs (e.g. QCI) and the Channel Quality Information (CQI) reported by the UEs to the eNBs.

Notice that the configuration of the scheduling policy in the eNBs from the SD-RAN Controller is not only done at the commissioning phase of the RAN slice, but can also be triggered at runtime, allowing in this way a dynamic adaptation of the policy applied per cell in order to compensate potential traffic unbalances across the overall, multi-cell scenario. This capability is supported through the so called *Scheduling Coordination Application* depicted in Fig. 3

IV. PROVISIONING OF RAN SLICES

The management of RAN slices in the testbed is structured in three phases: preparation, commissioning/de-commissioning and operation. The preparation phase includes the registration of the eNBs in the SDN-RAN controller and the initial configuration setup prior to the creation of the slices in the eNBs. Commissioning and de-commissioning actually refer to creation/termination of the RSIs, carrying out the necessary configurations and resource allocations over the affected eNBs. Finally, the operation includes the required actions to turn the RSI operational, i.e. serving UE's traffic. Each of the phases is described in detail in the following subsections.

*A. RAN Slice Preparation Phase*

Before provisioning RAN slices, the eNBs must be synchronized with the SD-RAN controller. The signalling exchanged during the synchronization process is depicted in Fig. 5 and is performed when an eNB running the 5G-EmPOWER Agent joins the network. To do this, the eNB must be registered at the 5G-EmPOWER OS. It should be noted that this task is carried out just once by the network administrator by introducing the eNB ID through the Web Service (GUI) in order to set it as a reliable 5G-EmPOWER-managed eNB.

After registration, the eNBs communicate their presence to the 5G-EmPOWER OS through the OpenEmpower protocol. This action is performed via the *Hello Request* message. In Fig. 5 it can be seen how this request is repeated as a "heartbeat" message while the connection is maintained. Upon receiving it, the OS replies using a *Hello Response* message, and after that, it sends to the eNB a *Capabilities Request* message in order to retrieve the operational information from this specific eNB that is relevant to the controller. This information, allocated in the *Capabilities Response* message from the eNB, includes data such as the eNB ID, and the operational settings of the active cells (e.g. cell identifiers, channel numbers, channel bandwidth). Based on this, the 5G-EmPOWER OS can build and update the network-wide RAN map. In addition to this, the *Capabilities Response* also allows the eNB to inform the 5G-EmPOWER OS about whether it supports RAN slicing capabilities. Otherwise, the eNB is not a valid node to deploy RAN slices.

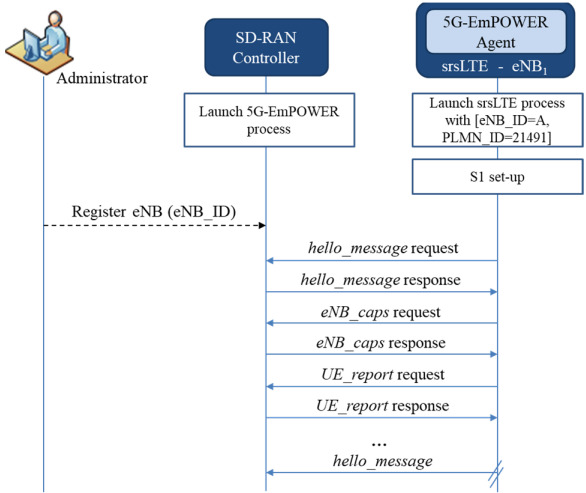

Fig. 5. Preparation phase (before a RAN slice can be created)

Finally, the OS solicits information regarding the UEs currently connected to the eNB through the *UE Report Request* message. Upon this, the eNB replies with a *UE Report Response*. This message provides the number of UEs with an active RRC connection to a cell handled by the eNB, along with the following information for each of these UEs: (1) the Cell Identifier (Cell ID), which provides a unique identifier of the specific cell to which the UE is connected; (2) the Non-Access Network (NAS) identifiers that could be extracted at the eNB, such as the International or Temporary Mobile Subscriber ID (IMSI / TMSI) and the Public Land Mobile Network Identifier (PLMN ID), which identify, respectively, the subscriber and the serving core network; (3) the Radio Network Temporary Identifier (RNTI), which is the temporary identifier allocated to the UE within a 3GPP RAN, and (4) information about the active DRBs, including the QCI and ARP. Of note is that the NAS identifiers will also include the S-NSSAI for 5G NAS signalling, though this is not currently implemented in the testbed that relies on the available 4G NAS signalling.

### B. RAN Slice Commissioning and De-commissioning Phases

The commissioning phase refers to the process following the preparation phase to create a new RAN slice (RSI). In particular, the creation of the new RSI starts with a request through the GUI provided by the 5G-EmPOWER OS. At this point, the slice configuration can be selected. This is done by setting the desired values of the *L2* and *L3 slice descriptors* according to the format reported in Table 2. The SD-RAN controller then verifies (and authorizes) that the selected values for the descriptors are valid and, if so, it instructs the eNB (through the 5G-EmPOWER Agent) to create the RSI through an *add slice request* message that includes the id of the slice (RSI_ID) to be created, along with the selected descriptor configuration. Upon receiving this message, the eNB registers the new slice and allocates the radio resources (i.e. resource units per slice [%]) according to the provided configuration. After these steps, the slice is ready for the activation, which is part of the operation phase described in the following section. The commissioning process is depicted in Fig. 6. Finally, it has to be pointed out that since the SD-RAN experimental testbed is based on 4G technology, thus the EPC and UEs do not support slicing, a list of NAS identifiers (e.g. IMSI, TMSI) is set during the request process to be used by the eNB for the association of UEs with the corresponding RSI.

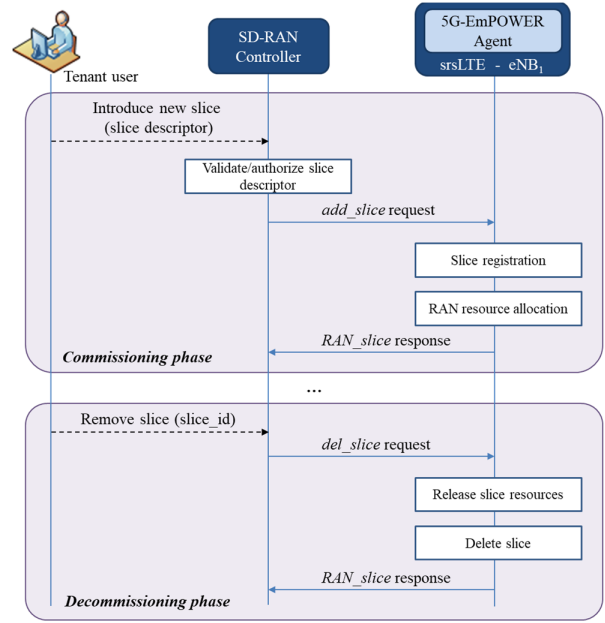

Fig. 6. Commissioning and decommissioning phases for the RAN slice

The lifecycle of a RAN network slice finishes with the decommissioning phase. At this point, the 5G-EmPOWER OS instructs the eNBs hosting the slice to release the resources allocated for serving its services. Notice that, from this moment the network slice is completely terminated and it is not available anymore. For that reason, in the case that any UE is attached to the slice, it will be automatically disconnected from that moment on. However, it should be noted that, upon a previous agreement between the network operators, these UEs could be migrated to another slice in order to maintain their services active. The communication performed between the SD-RAN controller and the eNB for the slice decommissioning is also shown in Fig. 6.

*C. RAN Slice Operation Phase*

This phase is composed of three major tasks: (i) activation; (ii) monitoring; and (iii) deactivation. The activation of the slice is carried out once the eNB has registered the slice and allocated the resources requested for such a slice. After this, the eNB (through the 5G-EmPOWER Agent) must provide the SD-RAN controller with a complete view of the active slices through the *RAN slice response* message. Upon the confirmation from the eNB, the slice is active and can be requested by any UE. After the activation, a monitoring function in the SD-RAN controller takes care of supervising the performance and resource utilisation of the slice (e.g. Key Performance Indicators (KPI) monitoring). In this phase, modifications of the RSIs can be triggered through the GUI, introducing changes in the slice descriptors. After validation/authorisation of the new configuration, the 5G-EmPOWER OS relies on the *RAN Slice Request*, a message that is sent to the eNBs providing an update in the slice, including information such as the AC policy, the UEs scheduling policy, and the percentage of resources assigned to a specific slice. Following this, the eNBs must apply the new configuration and, after the update, they must provide to the OS with an overview of the current status of the whole network through the *RAN slice response*. This message interchange is sketched in Fig. 7. Finally, the operation phase also considers the deactivation of a network slice instance, which could be later reactivated or decommissioned.

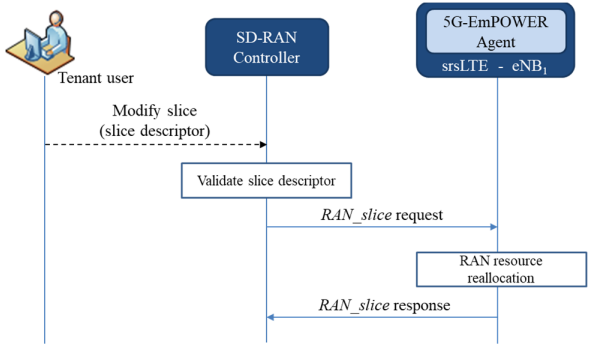

Fig. 7. Monitoring step within the RAN Slice operation phase

## V. TESTING OF RRM POLICIES FOR ADMISSION CONTROL

The operation of the SD-RAN testbed after the commissioning and activation of the RAN slices is validated through a testing scenario designed to showcase the operation of the implemented slice-aware AC function. For this purpose, a RAN slice (RSI_ID=1) is provisioned with the following AC policy: *L3Descriptor.Number of DRBs[(\*,\*)][cell][min] = 1* and *L3Descriptor.Number of DRBs[(\*,\*)][slice][max] = 2,* with no differentiation per QCI and ARP. This AC policy results in a minimum capacity guarantee of one DRB per cell, regardless of the number of DRBs activated in the other cells of the slice. On top of that, if the traffic demand in one cell exceeds such minimum capacity, a maximum number of 2 DRBs is enforced for the whole set of cells serving the slice. Let us notice, that these values have been chosen intentionally in order to force the UE rejection with the minimum number of devices as possible. For testing such a configuration, three commercial UEs (with test SIMs provisioned in the EPC) are switched on sequentially, get Internet connectivity through the SD-RAN testbed and start a video streaming application with downlink (DL) data rates of up to 3Mb/s. The testbed uses a channel of 10 MHz in the 2.6 GHz band, with PLMN ID=21491. During the connection of the UEs, the operation of the testbed is monitored, analysing the interactions between the testbed components and extracting specific details about the operation of the implemented AC function at both the eNB and the SD-RAN controller.

Figures 8, 9 and 10 showcase the signalling messages exchanged between UEs, eNB, SD-RAN Controller and EPC. For the sake of clarity, legacy 3GPP signalling is depicted with simple black lines, while the OpenEmpower protocol signalling is represented with blue bold lines. Moreover, the NAS messages, part of 3GPP signalling that are transparently transferred between UEs and EPC over RRC and S1-AP messages, are highlighted in green. All this information is collected from debugging and tracing capabilities that have been embedded within the eNB code.

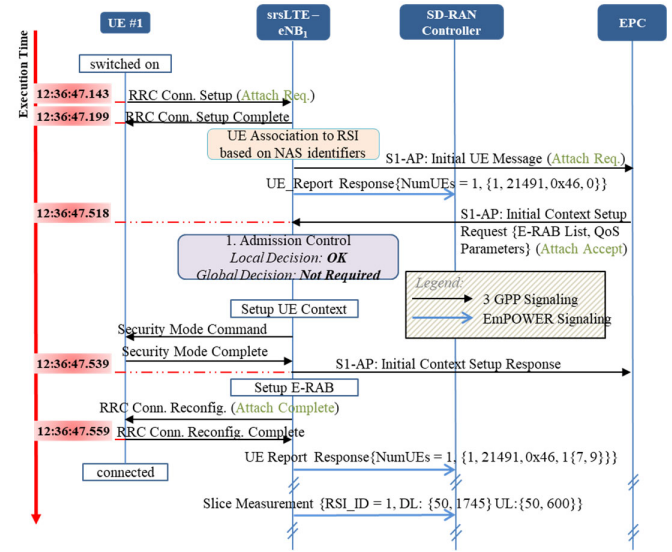

Fig. 8. Admission Control Process for UE#1

As it can be seen from Fig. 8, when UE#1 is switched on, a RRC connection is first established between the UE and the eNB. This triggers an *Initial UE Message* that embeds an *Attach Request* message originated in UE#1 from the eNB to the EPC, as well as a *UE Report Response* message from the eNB to the SD-RAN controller to notify about the presence of the new UE, which has been assigned the RNTI = 0x46. Note that at this point association between the UE and the RSI is solved by the eNB from the parameters (e.g. NAS identifiers) provided during the configuration of the RSI. As UE#1 is properly provisioned in the EPC database, an *Initial Context SetUp Request* is triggered to proceed with the creation of a UE Context in the eNB and the establishment of the E-RAB for data plane connectivity. At this point, the AC function within the eNB is executed (see (1) in Fig. 8) based on the RRMPolicy attributes (L3 slice descriptors) configured for the RSI_ID=1. In this case, since the minimum capacity guarantee is not exceeded (L3Descriptor.Number of DRBs[(*,*)][cell][min]=1) as no DRBs are activated yet for RSI_ID = 1, there is no need for interaction with the SD-RAN Controller. After the acceptance, the next action is to setup the UE context, activate the AS security, notify the EPC that the context has been setup successfully, internally configure the E-RAB within the eNB and finally send a *RRC Connection Reconfiguration* to UE#1 for DRB activation on the UE side. Finally, after the activation of the DRB (Setup E-RAB in Fig. 8), a new *UE Report Response* is sent by the eNB to the SD-RAN Controller with information about the QoS parameters of the established DRB (i.e. QCI=7 and ARP=9 in this case). As illustrated in Fig. 8, the duration of the overall process since the reception of the *RRCConnectionRequest* to the reception of the *RRCReconfigurationComplete* by the eNB takes around 416 ms. Also of note is that a *UE Report Response* message is sent by the eNB to the SD-RAN Controller each time a new RRC connection is turned up or down, while a *Slice Measurements* message, reporting the DL/UL PRB usage per slice is sent periodically. More particularly, the *Slice Measurements* message includes the RSI_ID, the DL/UL PRBs assigned to that slice, the used DL/UL PRBs during the measurement interval (i.e. 1 second) and other statistical information (not shown in the figures for clarity purposes). As depicted in Fig. 8, the *Slice Measurements* message reports a downlink occupation of 1745 PRBs in one second, which corresponds to a 17.45% radio load due to the video streaming application (note that 50 PRBs are available in the 10 MHz cell).

When UE #2 is switched on, an identical process is followed till the triggering of the AC function in the eNB (see (2) in Fig. 9). However, an interaction with the SD-RAN Controller is triggered in this case since the minimum capacity guarantee per cell has been reached with the previous activation of the UE#1 DRB in RSI_ID=1. Therefore, the admission decision depends now also on the total number of active DRBs within the slice (*L3Descriptor.Number of DRBs[(*,*)][slice][max] = 2*), which is known at controller level. For this reason, after the local decision made at the eNB, an *AC Request* message is sent to the SD-RAN Controller to trigger the centralized decision. This message includes the user RNTI (i.e. 0x47) and the details of the requested DRB (i.e. 1{7, 9} meaning 1 DRB with QCI=7 and ARP=9). Upon the reception of this message, the SD-RAN Controller checks the *L3Descriptor.Number of DRBs[(*,*)][slice][min] = 2* descriptor in order to accept or deny the DRB activation. Since in this case the number of active DRBs for RSI_ID=1 is equal to 1, it accepts the E-RAB establishment and sends the result (True) to the eNB with an *AC Response* message. The latter includes the user RNTI and the AC result. When the (positive) response is received by the eNB, the UE context is created, the EPC is notified, the E-RAB is established and finally the *RRC Conn Reconfiguration* is sent to the UE. In this case, the whole process takes around 575 ms due to the additional message exchange for the centralized AC control. Also of note is that now the resource utilization reported in the *Slice Measurements* message has increased to 25.5 % due to the traffic of the second user (UE#2).

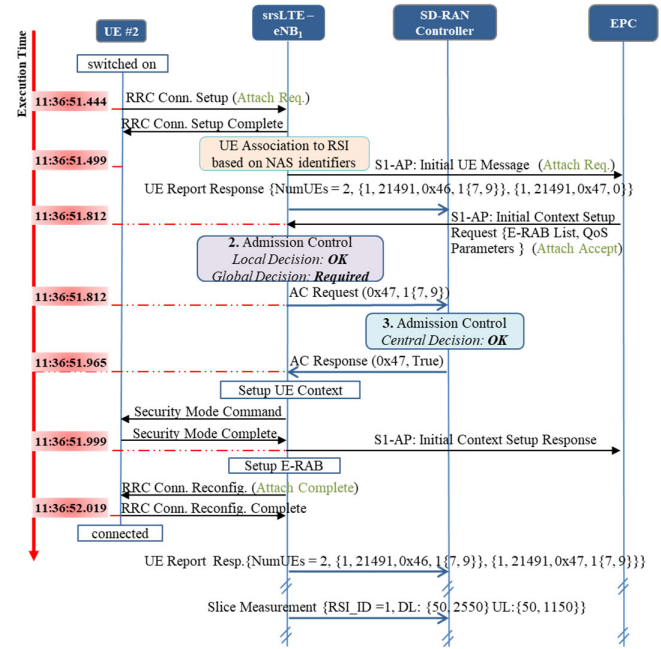

Fig. 9. Admission Control Process for UE#2

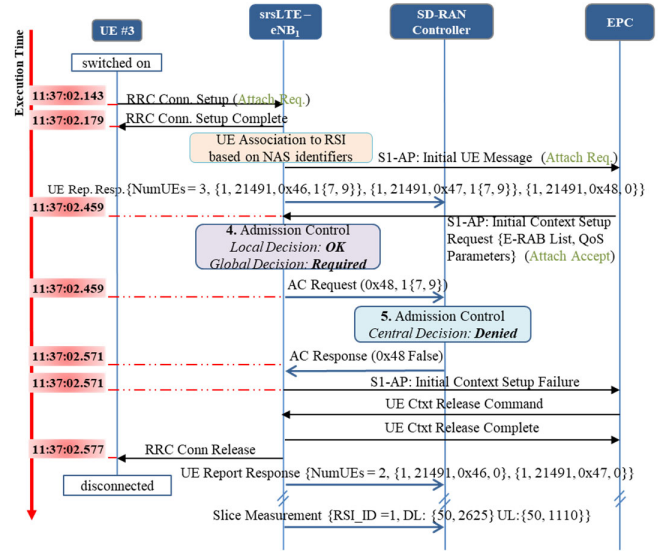

Fig. 10. Admission Control Process for UE#3

Finally, when UE #3 tries to attach to the slice, a similar process is followed (depicted in Fig. 10) as for the case of UE #2, with the difference that here the connection is denied through the centralized decision since the maximum number of active DRBs per slice has been already reached. Consequently, an *Initial Context SetUp Failure* is sent to the EPC (instead of sending an *Initial Context SetUp Response*) and the process is terminated with the *Context/Connection Release* message exchange between the EPC, eNB and UE. In this last case, the execution time of the whole process is 434 ms, from which 112 ms are due to the *AC Request*/*AC Response* message exchange. Since connection of

UE#3 was rejected, the resource utilization reflects the traffic of only UE#1 and UE#2.

The same test has been repeated 10 times following the same scenario setting and monitoring the message exchange delay in the eNB side. The average execution times measured for the establishment of the RRC connection and for completing the whole network registration setup, are given in Table 3, measuring separately the cases where AC decisions are made locally in the eNB and when centralized decisions in the SD-RAN Controller are also triggered (excluding the time required for the local decision). Moreover, the performance in the case of not using slicing is given as a benchmark.

As depicted in the table, similar execution times, of the order of 50 ms, are measured for the RRC connection setup in all the cases, since the slicing operation does not come with any additional processing to be carried during this procedure (differences in the average values observed between the three cases are only due to statistical fluctuations with the 10 measurements per case). On the other hand, for the whole network registration setup, it can be seen that centralized decisions lead to an increment in the execution time of the order of 200 ms. However, through the proper execution of each of the experiments it has been demonstrated that the addition of the slice-aware AC solution with distributed decision handling, presented in this work, is a feasible approach to follow and does not require any modification(s) to the standard settings of commercial UEs to cope with the extra time needed for the interaction between the eNB and the SD-RAN Controller.

|  | No slicing support (benchmark) | | Slice-aware RAN (Local AC decisions) | | Slice-aware RAN (Centralized AC decisions) | |
| --- | --- | --- | --- | --- | --- | --- |
|  | Avg (ms) | St. Dev. (ms) | Avg (ms) | St. Dev. (ms) | Avg (ms) | St. Dev. (ms) |
| RRC Connection SetUp | 52 | 6.8 | 50 | 5.4 | 44 | 8.9 |
| Whole network registration SetUp | 410 | 39.1 | 428 | 20.8 | 619 | 79.2 |

Table 3: Impact on the performance of the LTE service

## VI. CONCLUSIONS

This paper has presented a SD-RAN testbed that proves the feasibility and showcases the operation of (1) management services for the provisioning of RAN slices and (2) slice-aware/multi-cell scope RRM functions used to split the radio resources between RAN slices based on configurable RRM policy descriptors. The functional framework guiding the design of the testbed is in line with the latest 3GPP Release 15 specifications for network slicing management. In this respect, as a plausible realization of the *RRMPolicy* attribute included in the 3GPP information models, a template with a set of L3 and L2 descriptors has been proposed in this paper for a fine-grained specification of the RRM policy per slice at both cell and multi-cell levels. The testbed has been built leveraging open-source RAN distributions and the 5G-EmPOWER OS, which is the core component of the SD-RAN Controller providing the RAN slicing management functions. The main procedures and signalling exchanges supporting the preparation, commissioning and operation phases for RAN slicing provisioning have been showcased. Moreover, the operation of the slice-aware RRM functions for admission control have been analysed in detail in a testing scenario with commercial UEs. The achieved experimental results have verified the correct operation of the proposed solution and more importantly, it has been demonstrated that the operation of the commercial UEs is not affected by the extra time needed for the interaction between the eNB and the SD-RAN Controller.

Future work is envisaged to implement RRM algorithms able to exploit the full potential of the RRM policy descriptors proposed in this paper and test their operation under more complex scenarios with multiple cells and a mix of applications demanding diverse QoS characteristics. Moreover, testing of more sophisticated RRM algorithms (e.g. [26]) can be pursued.

ACKNOWLEDGEMENT

This work has been supported by the EU funded H2020 5G-PPP project 5G ESSENCE under the grant agreement 761592 and by the Spanish Research Council and FEDER funds under SONAR 5G grant (ref. TEC2017-82651-R).